\documentclass{rmaa}
\newcommand{\Msun}{\mbox{$\rm M_{\odot}$}}
\newcommand{\Rsun}{\mbox{$\rm R_{\odot}$}}

\usepackage{paralist}
\usepackage{graphicx}
\usepackage{psfrag,color}
\title{V1135 \,Herculis: a double-lined eclipsing binary with an Anomalous Cepheid \thanks{Based on 
  observations collected at T\"{U}B\.ITAK National Observatory (Antalya, Turkey).}}

\author{
E. Sipahi,\altaffilmark{1} 
C. \.{I}bano\v{g}lu,\altaffilmark{1}
 \"{O}. \c{C}ak{\i}rl{\i},\altaffilmark{1}
 H.A. Dal\altaffilmark{1}
 and S. Evren \altaffilmark{1}}

\altaffiltext{1}{Ege University, Science Faculty, Department of Astronomy and Space Sciences, 35100 Bornova, \.{I}zmir, TURKEY}

\shortauthor{Sipahi et al.}
\shorttitle{V1135 Her}
\fulladdresses{
\item Ege University, Science Faculty, Astronomy and Space Sciences Dept., 35100 
Bornova, \.{I}zmir, Turkey. $e-mail$: esin.sipahi@mail.ege.edu.tr}

\listofauthors{E. Sipahi, C. \.{I}bano\v{g}lu, \"{O}. \c{C}ak{\i}rl{\i}, H.A. Dal and S. Evren}
\indexauthor{Sipahi, E.}
\indexauthor{\.{I}bano\v{g}lu, C.}
\indexauthor{\c{C}ak{\i}rl{\i}, \"{O}}
\indexauthor{H.A. Dal}
\indexauthor{S. Evren}

\abstract{BVR light curves and radial velocities for the double-lined eclipsing binary V1135\,Her were obtained. The brighter component of V1135\,Her is a Cepheid variable with a pulsation period of 4.22433$\pm$0.00026 days. The orbital period of the system is about 39.99782$\pm$0.00233 days, which is the shortest value among the known Type\,II Cepheid binaries. The observed B, V, and R magnitudes were cleaned for the intrinsic variations of the primary star. The remaining light curves, consisting of eclipses and proximity effects, are obtained. Our analyses of the multi-colour light curves  and radial velocities led to the determination of fundamental stellar properties of both components of the interesting system V1135\,Her. The system consists of two evolved stars, G1+K3 between giants and supergiants, with masses of M$_1$=1.461$\pm$0.054 \Msun ~and M$_2$=0.504$\pm$0.040 {\Msun} and radii of R$_1$=27.1$\pm$0.4 {\Rsun} and R$_2$=10.4$\pm$0.2 {\Rsun}. The pulsating star is almost filling its corresponding Roche lobe which indicates the possibility of mass loss or transfer having taken place.  We find an average distance of d=7500$\pm$450 pc using the BVR magnitudes and also the V-band extinction. Location in the Galaxy and the distance to the galactic plane with an amount of 1300 pc indicate that it probably belongs to the thick-disk population. Most of the observed and calculated parameters of the V1135\,Her and its location on the color-magnitude and period-luminosity diagrams lead to a classification 
of an {\it Anomalous Cepheid}.    
}
\resumen{}
\addkeyword{ stars: binaries: close}
\addkeyword{ binaries: eclipsing}
\addkeyword{ binaries: general}
\addkeyword{ binaries: spectroscopic}
\addkeyword{ stars: individual: V1135\,Her}
\begin{document}
\maketitle

\section{Introduction}                                                                                                               \label{sec:intro}
Double-lined eclipsing binaries are known as astrophysical laboratories. Accurate fundamental physical properties of the stars, such as mass, radius and effective temperature, are determined directly from their observations. In addition they allow us to study internal structure, core or surface convection and evolution of the stars.  The majority of all kind of stars occur in binary systems. Eclipsing binary systems include $\delta$\,Scuti stars \citep{soy06}, $\gamma$\,Doradus variables \citep{iba07}, RR\,Lyrae variables \citep{prs08}, Population\,I Cepheids \citep{pie11}, Population\,II\, Cepheids \citep{wal02} and Anomalous Cepheids \citep{sip13} which represent an important testing basis for theories on stellar structure and evolution as well as pulsation mechanisms. Although these variables are located in a restricted region in the Hetrzsprung-Russell (HR) diagram, they have very different masses, effective temperatures and chemical abundances.  In general it is assumed that the components of a binary system are formed simultaneously from the parent cloud. Therefore the components should have the same chemical composition at the beginning of their main-sequence evolution. One can utilise information from the non-pulsating companion in identifying stellar evolution models for pulsating components. 

There are some metal-poor variables with periods between 0.5 and 3 or more days lying above the RR\,Lyrae-type stars which may have entirely different origin as compared to the normal Type\,II Cepheids in the globular clusters. These variables are called as Anomalous Cepheids (hereafter ACs). They are found in the general field, globular clusters and nearby dwarf spheroidal galaxies. Anomalous Cepheids are brighter than the Type\,II Cepheids a given color, and hence they follow a different P-L relation \citep{sos08b}. \citet{fio06} suggested that the ACs are extension of the Type\,I Cepheids to lower metallicities and masses.  Moreover, they predict a mass range for the ACs as 1.9$<$M/M$_{\odot}<$3, which may be a product of a mass about  M$<$4M$_{\odot}$, experienced the helium flash in the past.

V1135\,Her (NSV\,10993, 2MASS\,J18321299+1217042, GSC\,01032-01378, V=12$^{m}$.56, B-V=0$^{m}$.848)  was appeared as a 
variable star in the list of \citet{hof49} (see also \citet {kin00}). The light variability of V1135\,Her has been attributed to an eclipsing binary of a W\,UMa type by \citet{hof49} and \citet{goe56}. The orbital period of the eclipsing pair was estimated by \citet{ote05} as about 40 days and the system classified as an EA type with an amplitude of 0.40 mag while it was classified as an EW type in the NSV catalogue. Later on it was called as V1135\,Her and classified as an EB type eclipsing binary according to the rules of the GCVS \citep{kaz08}. \citet{khr08} detected for the first time pulsation in the eclipsing binary system V1135\,Her based on the NSVS publicly available data \citep{woz04}. He estimated a pulsation period of about 4.2243 days with an amplitude of about 0.2\, mag in the R-passband. However, he noted that the ASAS-3 data \citep{poj05} did not contradict his interpretation. V1135\,Her is an eclipsing binary with a Cepheid component an analogue of  TYC\,1031\,1262\,1, very recently interpreted by \citet{sip13}. It has the shortest orbital period among the eclipsing binaries having 
pulsating components. 
  
In this study we present our multi-color photometric and spectroscopic observations of V1135\,Her.  The main 
aim of this study is to derive the masses and radii of the components by analysing the light and radial velocity 
curves using the contemporary methods. Thus, mass and radius of a Cepheid variable will be revealed directly 
from the spectroscopic and photometric observations. We will also discuss the pulsation characteristics and its 
place in our Galaxy.

\section{Observations }
\subsection{Photometric observations}
The photometric observations in the wide-band Johnson BVR system were performed with the 35\,cm MEADE\,LX200\,GPS telescope at Ege University 
Observatory. The details of the systems can be found in  \citet{sip13}.  The BVR observations in 2008 were obtained on 47 nights between May\,30 and October\,22.  The observations in 2009 were obtained on 17 nights between July\,21 and October\,8. GSC\,1032\,795 and GSC\,1032\,1159 are taken as the comparison and check stars, respectively. Some basic parameters of the comparison stars are listed in Table\,1. Although the programme and comparison stars are very close in the sky, differential atmospheric extinction corrections were applied. The atmospheric extinction coefficients were obtained from observations of the comparison stars on each night. Moreover, the comparison stars were 
observed with the standard stars in their vicinity and reduced differential magnitudes, in the sense variable minus comparison, were transformed to the standard system. The standard stars are chosen from the lists of Landolt (1983, 1992). Heliocentric corrections were also applied to the times of the observations. The V-bandpass magnitude of the system is 12.56 which is not so bright for a 35 cm telescope. It was observed only three or more times in some nights. Therefore we consider that our photometric observations were affected from variable atmospheric conditions and position of the star on the sky.

In Fig.\,1 we plot the B-, V- and R-passband observations versus the pulsation period of the Cepheid variable. As it is seen the dominant light variations are originated from the intrinsic variations of the brighter component. The below shifted observations correspond to the eclipses. The standard deviations of each data point are about 0.025, 0.020, and 0.012 mag in B, V and R passbands, respectively. The observational 
data can be obtained from the authors.

\begin{figure*}
\center
\includegraphics[width=10cm,angle=0]{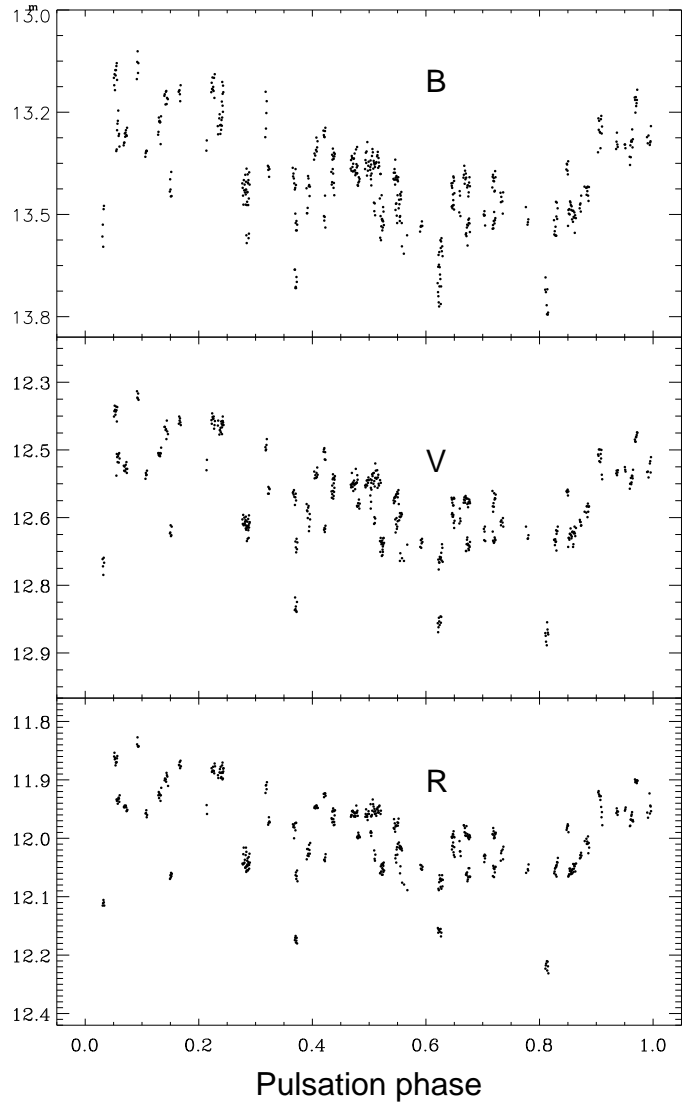}\\
\caption{The B-, V-, and R-passband light curves for V1135 \,Her. The ordinates and apsis are the standard magnitudes and the pulsation phases.} 
\end{figure*} 

\begin{table}
\centering
\footnotesize
\begin{minipage} {12 cm}
\caption{The coordinates, apparent visual magnitudes and the colors of the stars observed.}
\begin{tabular}{|l|c|c| c| c|}
\hline
\hline
Star	& $\alpha$ 	&$\delta$    &V (mag)	&	B-V(mag)	\\
\hline					
V1135\,Her       & 18$^h$ 32$^m$ 13$^s$	& 12$^{\circ}$ 17$^{'}$ 04$^{''}$.8     & 12.563	& 0.848		\\
GSC\,1032\,795 	 & 18 32 12  & 12 11 41     &  11.996	&  0.812  \\
GSC\,1032\,1159	 & 18 32 16  & 12 10 34     &  11.730   &  0.935	\\
\hline
\end{tabular}
\end{minipage}
\end{table}
\smallskip

\subsection{Spectroscopic observations}
Optical spectroscopic observations of V1135\,Her were obtained with the Turkish Faint Object Spectrograph Camera (TFOSC) attached to the 1.5\,m telescope on 13\,nights between July\,4,\,2011 and August\,28,\,2012 under good seeing conditions. Further details on the telescope and the spectrograph can be found at http://www.tug.tubitak.gov.tr. The wavelength coverage of each spectrum was 4000-9000 \AA\,in 11\,orders, with a resolving power of $\lambda$/$\Delta \lambda$ $\sim$7\,000 at 6563 \AA~and an average signal-to-noise ratio (S/N) was $\sim$120. We also obtained a high S/N spectrum of the 35\,Cyg (F6\,Ib), 36\,Per (F4\,III), $\alpha$\,Lyr (A0\,V) and HD\,50692 (G0\,V) 
for use as templates in derivation of the radial velocities.

The electronic bias was removed from each image and we used the 'crreject' option for cosmic ray removal. Thus, the resulting spectra were largely cleaned from the cosmic rays. The echelle spectra were extracted and wavelengths were calibrated by using Fe-Ar lamp source with help of the IRAF {\sc echelle} \citep{ton79} package. 
 
The stability of the instrument was checked by cross-correlating the spectra of the standard star against each other using the {\sc fxcor} task in IRAF. The standard deviation of the differences between the velocities measured using {\sc fxcor} and the velocities in \citet{nid02} was about 1.1 $kms^{-1}$.

\section{Pulsation and orbital periods}
\subsection{Pulsation period}
The pulsation period was estimated by \citet{khr08} as 4.2243\,d. Since the light variation of the pulsating component dominates in the light curve we first attempted to refine the pulsation period using the program PERIOD04 \citep{len05}. A Fourier power spectrum of all the available V-passband data gave a spectral peak at a frequency about $f_0$=0.2367 c/d which corresponds to 4.2248\,d. This package computes 
amplitudes and phases of the dominant frequency as well as simultaneous multi-frequency $sine$ wave fitting. Since the observations were obtained with different instruments in different years under dissimilar observing conditions, many systematic observational errors and computational errors affected the data. Therefore, we did not attempt to search for the second or the third frequencies for pulsation if they 
really do exist. As it will be explained in \S 4 we represented all available V-data, subtracting the eclipses, with a truncated $Fourier$ series which includes $cosine$ and $sine$  terms up to second order. We then  calculated the light variation originated from the oscillations of the more luminous component. We separated all the data with an interval of about 6\,days and determined maximum times by shifting the calculated light curve along the time axis. It should be noted here that the shape of the light curve is assumed to be more or less constant during the time base of the observations.  After the best fit is obtained, the times for the mid-maximum light are read off directly from the observations and presented in Table\,2. Although the observations obtained by NSVS and ASAS have relatively large scatters we had to use all the data because of limited observations both in time elapsed and the continuous observations due to its relatively longer pulsation period. From the time of maximum light and pulsating period given by \citet{khr08} we shifted the epoch and  using the following elements

\begin{equation}
Max (HJD)=2\,451\,265.430+4^d.2243\times E 
\end{equation}
we obtained the residuals between the observed and calculated times of mid-maximum light as well as the number of the elapsed cycles. In Fig.\,2 we plot the residuals O-C(I) versus the epoch numbers. The residuals are slightly increasing as the pulsation cycle is growing which indicates a correction of the pulsation period is needed. A least squares solution gives the following ephemeris,

\begin{equation}
Max (HJD)=2\,454\,653.979(0.148)+4^d.22433(0.00026) \times E 
\end{equation}
The standard deviations are given in the parentheses. The pulsation period of V1135 \,Her seems to constant during the observations. The pulsation period of the Cepheid is nearly the same with that estimated by \citet{khr08}.

\begin{figure*}
\center
\includegraphics[width=10cm,angle=0]{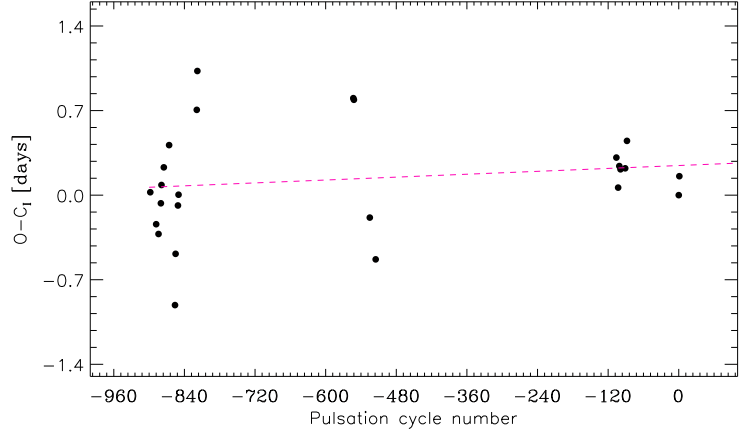}\\
\caption{The O-C residuals obtained by Eq.1 for the pulsating star of V1135 \,Her are plotted against the pulsation cycle number. A linear fit to the data is shown by dashed line.} 
\end{figure*}

\begin{table}
\caption{The times of mid-maximum light for V1135 \,Her. The O-C(I) and O-C(II) residuals were computed with Eqs.1 and 2, respectively.}
\begin{tabular}{lccccc}
\hline
HJD-2 450 000	& E	&O-C(I)    &O-C(II) & Filter	& $ Ref.$	\\
\hline	
1278.573	&	-799&	 0.190	&	-0.167	&	R	&	1	\\
1320.551	&	-789&	-0.454	&	-0.432	&	R	&	1	\\
1337.368	&	-785&	-0.535	&	0.513	&	R	&	1	\\
1354.519	&	-781&	-0.281	&	-0.259	&	R	&	1	\\
1358.895	&	-780&	-0.130	&	-0.108	&	R	&	1	\\
1375.938	&	-776&	 0.016	&	0.038	&	R	&	1	\\
1414.141	&	-767&	 0.200	&	0.222	&	R	&	1	\\
1455.060	&	-758&	 0.099	&	1.010	&	R	&	1	\\
1459.708	&	-756&	-0.699	&	-0.678	&	R	&	1	\\
1477.006	&	-752&	-0.299	&	-0.278	&	R	&	1	\\
1481.319	&	-751&	-0.210	&	-0.189  &	R	&	1	\\
1615.087	&	-720&	0.492	&	0.513	&	V	&	2	\\
1619.633	&	-719&	0.814	&	0.834	&	V	&	2	\\
2736.735	&	-454&	0.589	&	0.602	&	V	&	2	\\
2740.946	&	-453&	0.576	&	0.588	&	V	&	2	\\
2854.027	&	-426&	-0.400	&	-0.388	&	V	&	2	\\
2895.924	&	-416&	-0.745	&	-0.734	&	V	&	2	\\
4624.506	&	  -7&	0.097	&	0.097	&	V	&	3	\\
4636.930	&	  -4&  -0.151	&	-0.151	&	V	&	3	\\
4645.557	&	  -2&	0.027	&	0.027	&	V	&	3	\\
4653.978	&	   0&	0.000   &	0.000	&	V	&	3	\\
4687.781    &	   8&	0.008	&	0.008	&	V	&	3	\\
4700.681	&	  11&	0.235	&	0.235	&	V	&	3	\\
5071.970	&	  99&	-0.214 	&	-0.217	&	V	&	3	\\
5076.351	&	 100&	-0.057	&	-0.060	&	V	&	3	\\
\hline
\end{tabular}
\begin{list}{}{}
\item[Ref:]{\small (1) NSVS, (2) ASAS, (3) This study}
\end{list}
\end{table}

\subsection{Orbital period}
The orbital period of V1135\,Her was estimated by \citet{khr08} as 40\,days. We subtracted intrinsic variations of the more luminous star from all the available data, thus, the remaining light variations were assumed to be originated from the eclipses and proximity effects. We revealed light variations due to the eclipses and proximity analysis with a preliminary analysis of these data. The shape of the primary eclipse was revealed and compared by the observations which fall in the ascending and descending branches of the eclipse. Comparing the computed light curve with the observations, we obtained 11 times for mid-eclipse and presented in Table\,3. The low precision is due to times of minima being based upon very few observations in most cases. The O-C(I) residuals were computed using the ephemeris

\begin{equation}
Min (HJD)=2\,453\,126.2583+39^d.99844\times E 
\end{equation}
where the orbital period was derived using the program PERIOD04. In Fig.\,3 we plot O-C(I) residuals versus the epoch numbers.  The observations cover a short time interval (about 10 years). Since there are four or more nights observations on decreasing and increasing parts of the 
primary minimum obtained on JD 2\,451286, 2\,452726, 2\,454646 and 2\,454686 a weight of 5 was taken. A weighted linear least
squares fit to the data gives the following ephemeris,

\begin{equation}
Min (HJD)=2\,453\,126.678(0.083)+39^d.99782(0.00233) \times E  
\end{equation}
The orbital period is very close to that estimated by \citet{khr08}. 

\begin{table}
\footnotesize
\begin{minipage} {12 cm}
\caption{The times of mid-minimum light for V1135 \,Her. The O-C(I) and O-C(II) residuals were computed with the 
ephemeris given by Eqs.\,3, and 4,  respectively.}
\begin{tabular}{lccccc}
\hline
HJD-2 450 000	& E	&O-C(I)    &O-C(II) & Filter	& $ Ref.$	\\
\hline	
1286.75&	-46	&	-0.46	&	-0.03	&	R	&	1	\\
1326.94&	-45	&	-0.26	&	 0.16	&	R	&	1	\\
1366.93&	-44	&	-0.26	&	 0.15	&	R	&	1	\\
2726.28&	-10	&	-0.51	&	-0.42	&	V	&	2	\\
3126.67&	0	&	 0.00	&	-0.00	&	V	&	2	\\
3486.48&	9	&	-0.09	&	-0.18	&	V	&	2	\\
4366.93&	31	&	 0.61  &	 0.32	&	V	&	2	\\
4646.76&	38	&	 0.52	&	 0.16	&	V	&	3	\\
4686.79&	39	&	 0.56	&	 0.19	&	V	&	3	\\
4726.11&	40	&	-0.10	&	-0.48	&	V	&	3	\\
5046.69&	48	&	 0.57	&	 0.12	&	V	&	3	\\
\hline
\end{tabular}
\end{minipage}
\begin{list}{}{}
\item[Ref:]{\small (1) NSVS, (2) ASAS (3) This study}
\end{list}
\end{table}

\begin{figure*}
\center
\includegraphics[width=10cm,angle=0]{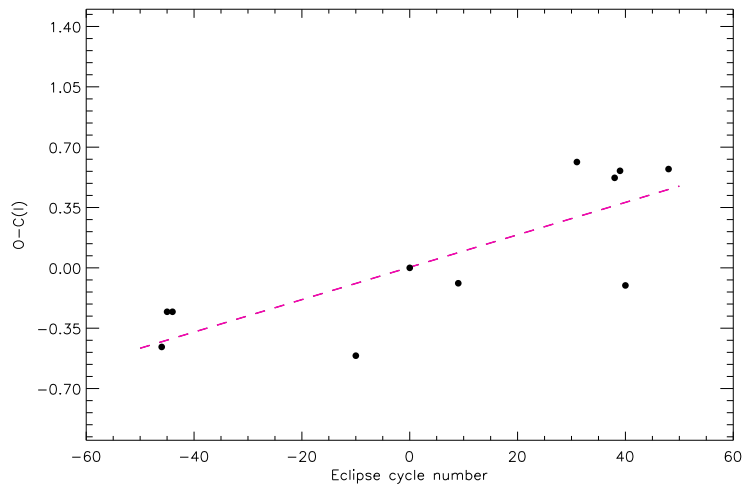}\\
\caption{The O-C(I) residuals obtained by Eq.3 for eclipsing binary V1135\,Her and a linear least squares fit to the data.} 
\end{figure*}

\section{Analysis}
\subsection{Effective temperature of the primary star}
We have used our spectra to reveal the spectral type of the primary component of V1135\,Her. For this purpose we have degraded the spectral resolution from 7\,000 to 3\,000 by convolving them with a Gaussian kernel of the appropriate width, and we have also measured the equivalent widths ($EW$) of the  photospheric absorption lines for the spectral classification. We have followed the procedures of  \citet{her04}, choosing helium lines in the blue-wavelength region, where the contribution of the secondary component to the observed spectrum is almost negligible. From several spectra we measured $EW_{\rm He I+ Fe I\lambda 4922 }=0.81\pm 0.07$\,\AA, $EW_{\rm He I+ Fe I\lambda 4144 }=0.33\pm 0.05$\,\AA, $EW_{\rm Ca I\lambda 5589 }=0.35\pm 0.07$\,\AA, $EW_{\rm Ca I\lambda 6162 }=0.35\pm 0.07$\,\AA, and $EW_{\rm CH (G-band) \lambda 4300}=1.05\pm 0.02$\,\AA. From the calibration relations $EW$--Spectral-type of \citet{her04}, we have derived a spectral type of G1\,II-III for the more luminous star with an uncertainty of about 1 spectral subclass. In Fig.\,4 we compare spectrum of the variable, obtained on JD\,24\,56136 with the spectra of some standard stars.

We have also observed the variable and comparison stars with the standard stars at the same nights. The average standard magnitudes and colors of the variable are obtained as $<$V$>$ =12.563$\pm$0.020, $<$B-V$>$=0.848$\pm$0.040 and $<$V-R$>$=0.551$\pm$0.040 mag. Comparing the location of the variable in the (B-V)-(V-R) diagram given by \citet{dri00} we estimate a spectral type of G1-II\,-III which is in a good agreement with that derived from spectroscopy. The observed infrared colors of J-H=0.316$\pm$0.029 and H-K=0.121$\pm$0.028 are obtained using the JHK magnitudes given in the 2MASS catalog \citep{cutri}. These colors correspond to a G0$\pm$2 supergiant star \citet{dri00} which is consistent with that estimated both from the spectra and BVR photometry. The effective temperature deduced from the calibrations of \citet{jag87} and \citet{dri00} is 5\,280$\pm$120\,K. The standard deviations are estimated from the spectral-type uncertainty. From the tables given by \citet{dri00} the intrinsic (B-V) color of 0.82 mag and bolometric correction of 0.20 mag were interpolated. Therefore, an interstellar reddening of E(B-V)=0.028\,mag is estimated.

\begin{figure*}
\center
\includegraphics[width=10cm,angle=0]{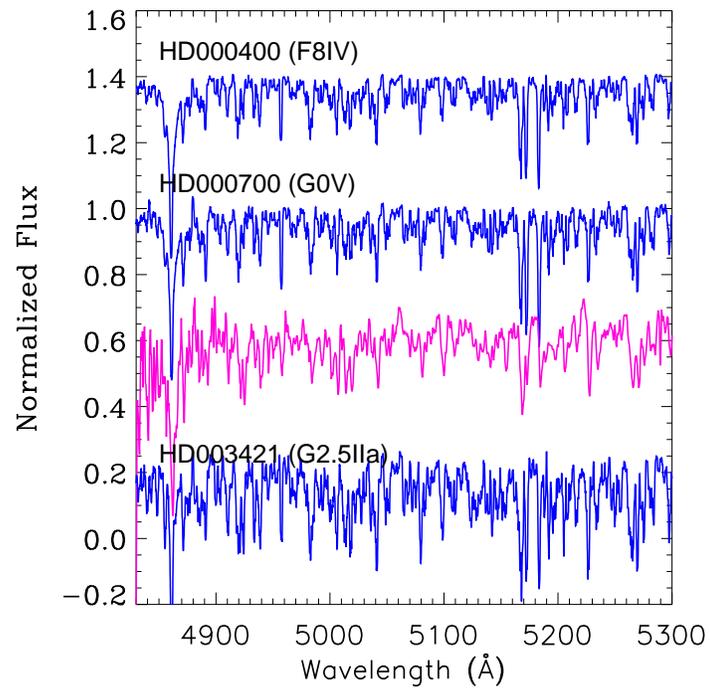}\\
\caption{Comparison of the spectrum of V1135\,Her with some standard stars with similar spectral type but different luminosity.} 
\end{figure*}

\subsection{Radial velocities}
To derive the radial velocities of the components, the 13\,TFOSC spectra of the eclipsing binary  were cross-correlated against the spectrum of standard stars, order-by-order basis using the {\sc fxcor} package in IRAF. The majority of the spectra showed two distinct cross-correlation 
peaks in the quadrature, one for each component of the binary. Thus, both peaks were fitted independently in the quadrature with a $Gaussian$ profile to measure the velocity and errors of the individual components. If the two peaks appear blended, a double Gaussian was applied to the combined profile using the {\it de-blend} function in the task. Here we used as weights the inverse of the variance of the radial velocity measurements in each order, as reported by {\sc fxcor}. 

\begin{table}
\caption{Heliocentric radial velocities of V1135 \,Her. The columns give the heliocentric Julian date, the orbital phase (according to the ephemeris in Eq.~3), the radial velocities of the two components with the corresponding standard deviations.}
\begin{tabular}{@{}ccccccccc@{}c}
\hline
HJD 2400000+ & Phase & \multicolumn{2}{c}{Star 1 }& \multicolumn{2}{c}{Star 2 } 	\\
             &       & $V_p$                      & $\sigma$                    & $V_s$   	& $\sigma$	\\
\hline
55747.3611	&0.5384  &-1.2	&4.3   &-19.9   &5.1\\
55751.3423	&0.6379  &8.4	&3.2   &-44.4   &4.3\\
55796.3308	&0.7630	 &15.0	&0.8   &-61.9   &4.1 \\
55800.4144	&0.8651	 &11.7	&3.1   &-49.1   &3.6 \\   
55835.3383	&0.7385	 &14.3	&1.1   &-58.9   &4.4\\ 
55864.2467	&0.4614	 &-9.1	&5.2   &11.2    &6.6 \\
56131.3862	&0.1418  &-17.1	&1.5   &40.1    &4.3\\
56133.4240	&0.1928  &-19.4	&1.7   &44.5    &3.4\\
56134.3924	&0.2170	 &-22.2	&1.3   &49.2    &3.2 \\
56135.4096	&0.2424	 &-18.9	&1.8   &50.9    &3.3 \\   
56135.4505	&0.2684	 &-19.9	&1.8   &51.1   &3.8\\ 
56137.3887	&0.2919  &-21.1	&2.0   &49.1    &3.4 \\
56168.2987	&0.0649  &-10.1	&6.5   &17.8    &7.7 \\
\hline \\
\end{tabular}
\end{table}

The heliocentric radial velocities for the primary (V$_p$) and the secondary (V$_s$) components are listed in Table\,4, along with the dates of observations and the corresponding orbital phases computed with the new ephemeris given in previous section. The uncertainties of the measured velocities are derived from different orders. Of course, the velocities of the primary star are slightly affected from its pulsation. However, we could not reveal the velocity contribution originated from the pulsation. Our resolution is inadequate to do this. We used HD 50692, Sp.G0V, as a primary standard in the cross-correlation which seems to adequate for deriving radial velocities of both components. The radial velocities are plotted against the orbital phase in Fig.\,5. The weight $W_i = 1/\sigma_i^2$ has been given to each measurement. The standard errors of the weighted means have been calculated on the basis of the errors ($\sigma_i$) in the velocity values for each order according to the usual formula. The $\sigma_i$ values are computed by {\sc fxcor} according to the fitted peak height, as described by \citet{ton79}.  Radial velocity changes of the Cepheid caused by the pulsation are ignored. We did not attempt to decomposition the radial velocity measurements of the primary star into the pulsation radial velocity and the orbital radial velocity. 

First we analysed the radial velocities for the initial orbital parameters. We used the orbital period held fixed and computed the eccentricity of the orbit, systemic velocity and semi-amplitudes of the radial velocities. The results of the analysis are as follows: $e$=0.001$\pm$0.001, i.e. formally consistent with a circular orbit, $V_{\gamma}$= -4.64$\pm$0.42 $kms^{-1}$, $K_1$=19.27$\pm$0.99 and $K_2$=55.88$\pm$0.57 $kms^{-1}$. Using 
these values we estimate the projected orbital semi-major axis and mass ratio as: $a$sin$i$=59.39$\pm$0.91 \Rsun~ and $q=\frac{M_2}{M_1}$=0.345$\pm$0.027.

\begin{figure*}
\center
\includegraphics[width=14cm,angle=0]{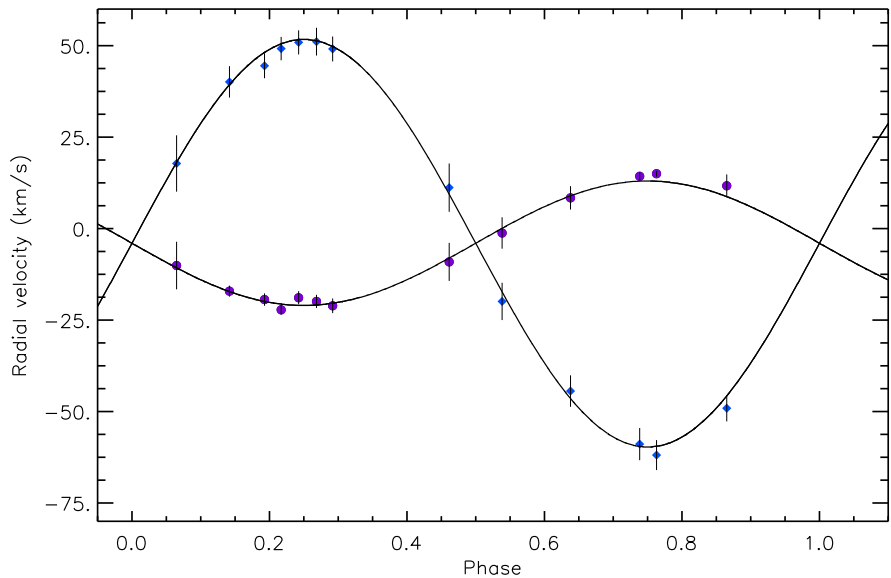}
\caption{Radial velocities folded on a period of 39.99782\,days and the model. Dots (primary) and squares (secondary) with error bars show the radial velocity measurements for the components of the system } 
\end{figure*}

\subsection{Intrinsic variations of the primary star}
The eclipsing  binaries provide critical information about the orbital parameters such as orbital inclination, fractional radii, luminosities, ratio of effective temperatures etc. If the eclipsing binary is a double-lined binary, the masses and radii of the component stars can be determined in solar units. Using the inverse-square law one can accurately determine distance to the system which is independent of all distance methods. The observed light variations are composed of intrinsic light variations of the more luminous star and mutual eclipses. First we subtracted all the observations within the eclipses. The remaining observations are phased with respect to the pulsation period of 4.22433\,days. The light curves of the Cepheid show slow decline and rapid rise, i.e., asymmetric light curve with a sharp maximum. The observed magnitudes, colours and rise times are given in Table\,5. The median magnitudes are taken as 13.347, 12.514 and 11.964 mag for B, V and R passbands, respectively. The observed magnitudes are transformed to the flux using the median magnitudes. Then, we represented the intrinsic light variations of the primary star with a truncated Fourier series. Trial-and-error method showed that the observed light curves can well be represented by the second-order Fourier series. A preliminary light curve analysis is obtained and the light variations out-of-eclipses, originated from the proximity effects, are revealed. These variations are subtracted from the original observations and light variations due to the pulsation are obtained. The truncated Fourier coefficients are given in Table\,6 and the fits are compared with the observations in Fig.\,6.  

\begin{table}
\scriptsize
\caption{The observed magnitudes, colours, amplitudes and rise times for the Cepheid variable. }
\setlength{\tabcolsep}{0.8pt} 
\begin{tabular}{lrrrr}
\hline
Passband  & { $max$} & $min $   & $min-max $   & $M-m $                \\
\hline	
B 								&13.181 				&13.464					&0.283	 &0.388    \\
V 								&12.371					&12.600					&0.229	 &0.378    \\
R 								&11.840					&12.030					&0.190	 &0.400    \\
B-V 							&0.810					&0.864					&0.054   &        \\
V-R 							&0.531					&0.570					&0.039   &        \\
\hline
\end{tabular}
\end{table}

\begin{figure*}
\includegraphics[width=12cm,angle=0]{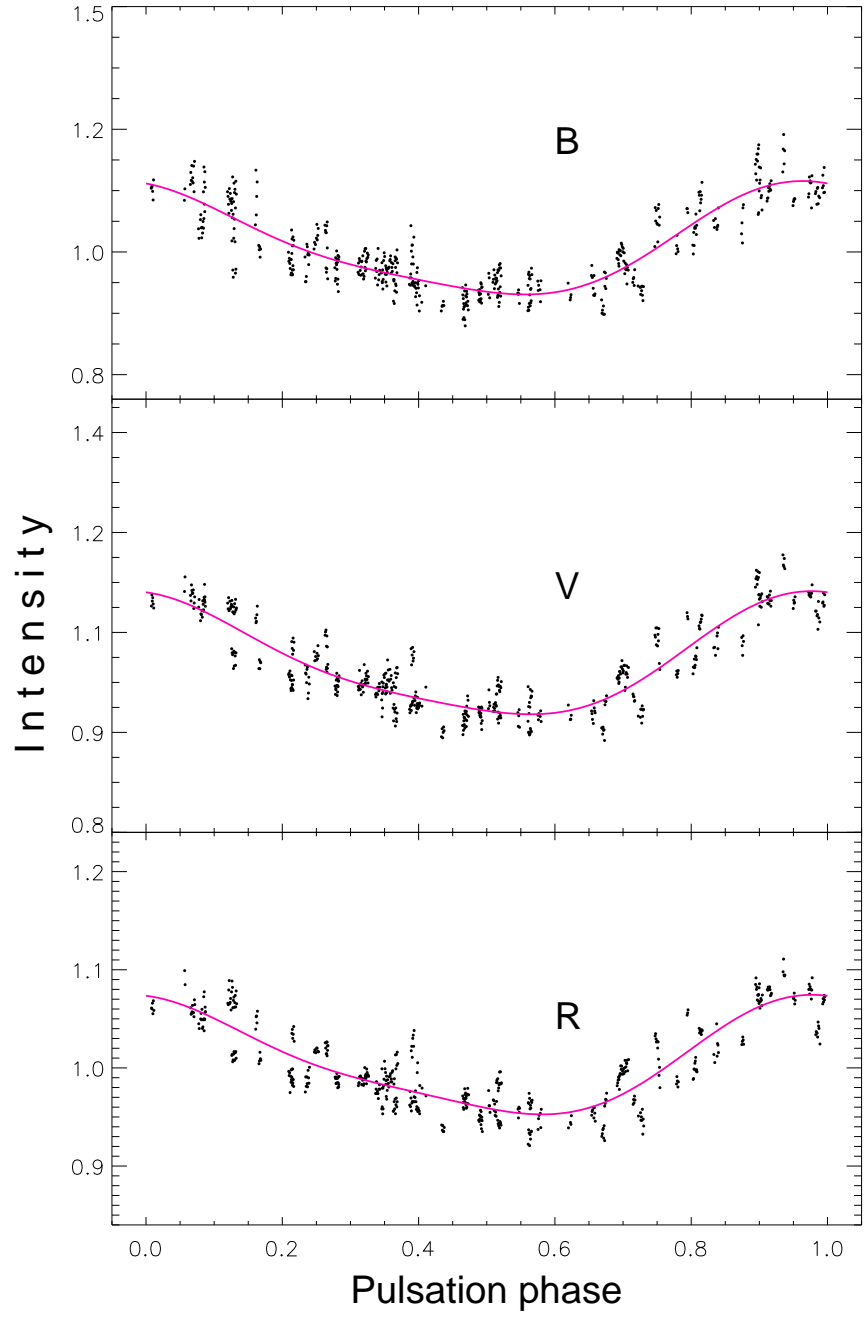}\\
\caption{ The B, V, and R passbands light curves of the pulsating primary star and their Fourier representations (solid lines) 
with the coefficients given in Table\,6. Note that the ordinates are normalized intensities. } 
\end{figure*}

\begin{table}
\scriptsize
\caption{Fourier coefficients of the oscillation light curves for V1135 \,Her. }
\setlength{\tabcolsep}{0.8pt} 
\begin{tabular}{lrrr}
\hline
Parameters  & { $B $} & $V $ & $R $                    \\
\hline	
A$_{0}$ 								&1.0156$\pm$0.0017 					&1.0081$\pm$0.0012			&1.0092$\pm$0.0011			    \\
A$_{1}$ 								&0.1103$\pm$0.0024					&0.0889$\pm$0.0018			&0.0715$\pm$0.0016			 \\
A$_{2}$ 								&0.0137$\pm$0.0023					&0.0132$\pm$0.0017			&0.0111$\pm$0.0015			    \\
B$_{1}$ 								&-0.0065$\pm$0.0024					&0.0005$\pm$0.0017			&0.0050$\pm$0.0016			 \\
B$_{2}$ 								&-0.0175$\pm$0.0024					&-0.0112$\pm$0.0017			&-0.0110$\pm$0.0016		     \\
\hline
\end{tabular}
\end{table}

\subsection{Analysis of the light curves }
The intrinsic light variations of the primary star were computed for each oscillation phase using the coefficients given in Table\,6. After subtraction of the Cepheid light changes from the observations we obtained light variations, consisting of only from the eclipses and proximity effects. In Fig.\,7 the eclipsing light curves in the B-, V- and R-passband are plotted versus the orbital phases calculated by the ephemeris given in Eq.\,4. The light curve of the system with curved maxima resembles to those $\beta$ Lyrae-type eclipsing binaries. 

We have used the most recent version of the eclipsing binary light curve modeling algorithm of \citet{wil71} (with updates, hereafter W-D), as implemented in the {\sc phoebe} code of \citet{prs05}. It uses the computed gravitational potential of each component to calculate the surface gravities and effective temperatures. The radiative characteristics of the stellar disks are determined using the theoretical Kurucz atmosphere models. The code needs some input parameters, which depend upon the physical properties of the component stars. The BVR photometric observations for the system were analysed simultaneously. We fixed some parameters whose values were estimated from spectra, such as effective temperature of the hotter component and mass-ratio of the system, which are the key parameters in the W-D code. The effective temperature of the primary star has already been derived from various spectral type-effective temperature calibrations as 5\,280 K and the mass-ratio from the semi-amplitudes of the radial velocity curves as 0.345. A preliminary estimate for the effective temperature of the cooler component is made using the depths of the eclipses in the BVR passbands. Therefore, initial linear and bolometric limb-darkening coefficients for the primary and secondary components were interpolated from \citet{vanham03}, taking into account the effective temperatures and the wavelengths of the observations. The bolometric albedos were adopted from \citet{lucy} as 0.5, typical for a fully convective stellar envelopes. The gravity-darkening exponents are assumed to be 0.32 for  both components, because the stars are cool and are assumed to have convective envelopes. The rotational velocities of the components are assumed to be synchronous with the orbital one and zero eccentricity. At first, we used Mode\,2 of the W-D code which is for detached binaries with  no constraints on the potentials. In this mode the fractional luminosity of the secondary component is computed from the other parameters via black body or stellar atmosphere assumption. 

\begin{figure*}
\center
\includegraphics[width=9cm,angle=0]{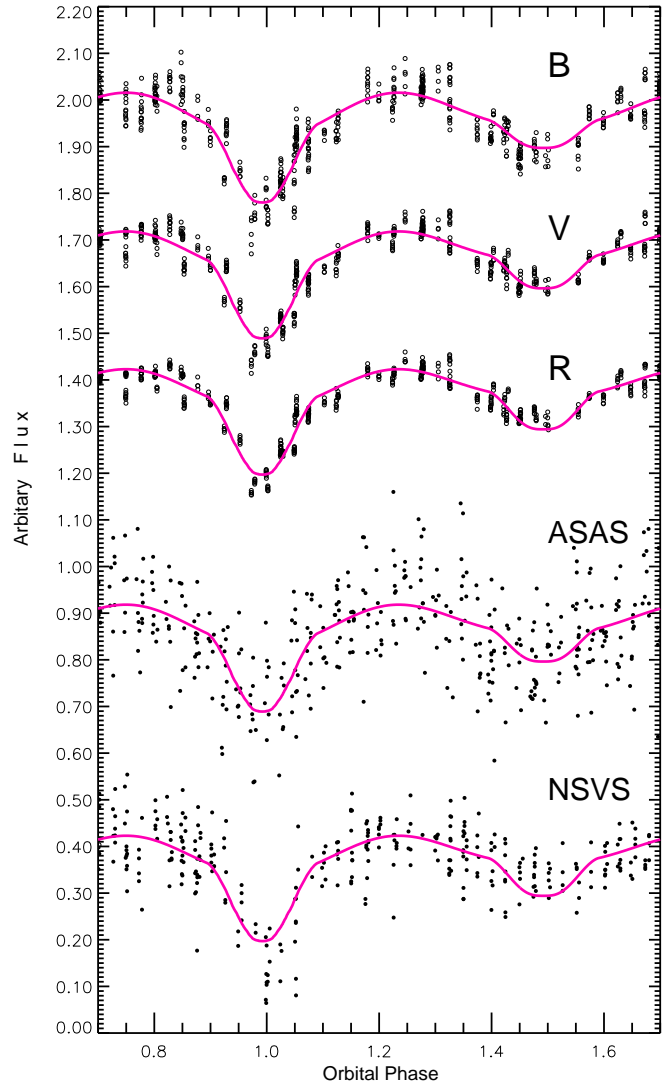}\\
\caption{The B, V and R passbands light curves, originated only from the eclipses and proximity effects, and the computed light curves. The solid lines are for the theoretical curves which are also compared by the ASAS V- and NSVS R-passband observations in the bottom two panels. }
 \end{figure*}

The adjustable parameters in the differential correction calculation are the orbital inclination, the dimensionless surface potentials, the effective temperature of secondary, and the monochromatic luminosity of the hotter star. Our final results are listed in Table\,7 and the computed light curves (continuous line) are compared with the observations in Fig.\,7.  In Table 7 the mean radii of the components are given. The Wilson-Devinney code calculates four dimensions of each component. We give only mean values of the radii which are used absolute values of the stellar radii.  The distortions on the shape for the primary and secondary are Rpole/Rpoint=0.852 and 0.967, respectively.  The shape distortion of more massive star is larger than that of the secondary. The uncertainties assigned to the adjusted parameters are the internal errors provided directly by the Wilson-Devinney code. In the last line of Table\,7 the sum of the squares of residuals ($\chi^2$), is also given. No reliable solutions could be obtained for Mode 4 for the semi-detached binaries with more massive star filling its Roche lobe and Mode\,6 in which both stars fill their limiting lobes. We could not obtained any convergent solution from the ASAS and NSVS data. The scatter of the observations are too large to be obtained a reliable orbital solution. In the bottom two panels of Fig.\,7 we compare our solution with the V- and R-passband observations of ASAS and NSVS, respectively.

We have also applied a procedure, a successive approximation method to prevent scatters in the light curves.  After subtraction light variation due to pulsation the remaining light curves were analyzed and the preliminary orbital elements were derived.  The light variation originated from the proximity effects removed from the original observations and light variations of the pulsating component were revealed. Then the procedure is repeated again. Unfortunately the scatters could not be removed.  

\begin{table}
\scriptsize
\caption{Results of the BVR light curves analyses for  V1135 \,Her. }
\setlength{\tabcolsep}{0.8pt} 
\begin{tabular}{lr}
\hline
Parameters  & { $BVR$}                   \\
\hline	
$i^{o}$										&74.56$\pm$0.18       \\
T$_{eff_1}$ (K)								 &5280[Fix]          \\
T$_{eff_2}$ (K)								&4250$\pm$50        \\
$\Omega_1$									&2.688$\pm$0.012    \\
$\Omega_2$									&3.368$\pm$0.033    \\
r$_1$				   						&0.4405$\pm$0.0025  \\
r$_2$										&0.1682$\pm$0.0027  \\
$\frac{L_{1}}{(L_{1}+L_{2})}$ 	            &0.973$\pm$0.004, 0.960$\pm$0.004, 0.950$\pm$0.004     \\
$\chi^2$									&1.470	              \\	
\hline
\end{tabular}
\end{table}

\section{Results and Discussion}
Combining the spectroscopic results along with the photometric solutions, listed in Table\,7, the absolute masses, radii, luminosities and surface gravities of the stars are obtained and presented in Table\,8. The mass and radius of the pulsating component are determined for the first time directly from radial velocities and multi-color light curves. They are unexpectedly large compared to the Type\,II Cepheids and small to the Type\,I Cepheids. The luminosity and absolute bolometric magnitude M$_{bol}$ of each star were computed from their effective temperatures and radii. The  bolometric magnitude and effective temperature for the Sun are taken as 4.74 mag and 5770\,K (\citet{dri00}). After applying bolometric corrections to the bolometric magnitudes, we obtained absolute visual magnitudes of the components. Taking into account the light contributions of the stars and total apparent visual magnitude we calculated their apparent visual magnitudes. The light contribution of the secondary star $L_2$/($L_1$+$L_2$)=0.027, 0.040 and 0.050 are obtained directly from the simultaneous B-, V-, and R-bandpass light curve analysis, respectively. This result indicates that the light contribution of the less massive component is very small, indicating its effect on the color at out-of-eclipse is almost negligible for the shorter wavelengths.

For the bolometric corrections given by \citet{dri00} we calculated a mean distance to V1135\,Her as $d$=7500$\pm$450 pc from BVR and JHK magnitudes. The distance from the galactic plane is calculated as 1300 pc using its galactic latitude of 9.78 degrees, at least 18 times larger than the mean scale height of the Type\,I Cepheids. The proper motion of this distant Cepheid are given by \citet{zac04} with too large 
errors. Therefore, we did not attempt to calculate Galactic velocity components.  However, its location in the Galaxy indicates that V1135\,Her is probably belonging to the thick-disk population. This result confirms the earlier suggestion of \citet{har84} that many field  Type\,II Cepheids are metal rich and have kinematic properties similar to the old-disk or thick-disk population.  

In  Fig.\,8 we plot known Type\,II Cepheids in the color-magnitude diagram. The binary Type\,II Cepheids taken from \citet{har84}, \citet{nem94}, and \citet{sip13} were also plotted on the M$_v$-(B-V) diagram. V1135 \,Her is located among the brightest Type\,II Cepheids close to the red edge of IS, with a luminosity of about 500 times solar. V1135 \,Her seems to an analogue of the binary Cepheids IX\,Cas with a pulsation period of 9.2 d, TX\,Del with a period of 6.2 d and AU\,Peg with a period of 2.4 d, ST\,Pup with a period of 18.5 d, and TYC\,1031\,1262\,1 with a period of 4.15 d. They share the limited region of the IS of Type\,II 
Cepheids. 

\begin{figure*}
\center
\includegraphics[width=12cm,angle=0]{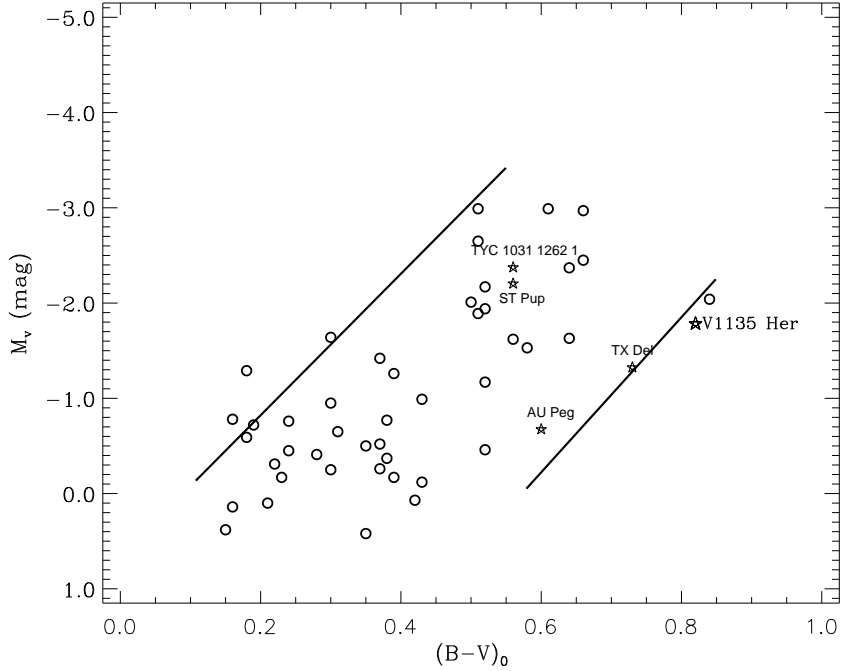}
\caption{The position of V1135 \,Her on the color-magnitude diagram. The circles refer to the Type\,II Cepheids in globular clusters and the stars for the binary Cepheids: V1135 \,Her, TYC\,1031\,1262\,1, ST Pup, AU Peg and TX Del. The lines show the borders of instability strip for Type\,II Cepheids.}
\end{figure*}

\begin{table}
 \setlength{\tabcolsep}{2.5pt} 
  \caption{Fundamental parameters of  V1135 \,Her.}
  \label{parameters}
  \begin{tabular}{lcc}
  \hline
  & \multicolumn{2}{c}{\hspace{0.25cm} V1135 \,Her} 			\\
   Parameter 												& Primary	&	Secondary					\\
   \hline
   Spectral Type											& G1($\pm$1)II-III  	&K3($\pm$1)II-III    				\\  
   Mass (M$_{\odot}$) 								& 1.461$\pm$0.054       &0.504$\pm$0.040			\\
   Radius (R$_{\odot}$) 								& 27.1$\pm$0.4    &10.4$\pm$0.2				\\
   $T_{eff}$ (K)											& 5\,280$\pm$120&4\,250$\pm$50    	\\
   Luminosity (L$_{\odot}$) 								& 517$\pm$50   & 32$\pm$4				\\
   Gravity ($cgs$) 										& 54$\pm$1          &129$\pm$8	                \\
   $(vsin~i)_{calc.}$ (km s$^{-1}$)			& 34.3$\pm$0.6	& 13.1$\pm$0.3		       							\\
   $a$ (R$_{\odot}$)									&\multicolumn{2}{c}{61.62$\pm$0.95}			\\
   $V_{\gamma}$ (km s$^{-1}$)				&\multicolumn{2}{c}{-4.64$\pm$0.42} 				     	\\
   $i$ ($^{\circ}$)										&\multicolumn{2}{c}{74.56$\pm$0.18	} 			\\
   $q$															&\multicolumn{2}{c}{0.345$\pm$0.018} 	\\   
   $d$ (pc)													& \multicolumn{2}{c}{7500$\pm$450}			\\
\hline  
 \end{tabular}
\end{table}

P-L relations for the Cepheids are studied by many authors (\citet{sto11}, \citet{stor11}, and \citet{sto12} and references therein). The P-L relation slope and zero points have long been discussed whether they are depended on the metallicity or not. \citet{fre01} suggest that the zero point is depended on the metal abundance in the stars. \citet{stor11} find identical Milky Way and LMC P-L relation slopes in the near-infrared bands by comparing LMC Cepheid P-L relations to their Milky\,Way counterparts. However, zero points show a slight metallicity effect, in the sense that metal-poor Cepheids are fainter than metal-rich Cepheids. The largest sample of ACs has been collected for the LMC by the OGLE III survey \citep{uda08}. \citet{sos08a} presented in the first part of the OGLE-III Catalog of Variable Stars data for 3361 Type\,I Cepheids in the LMC. Of which 1848 are fundamental-mode (F), 1228 first overtone (FO), 14 second overtone (2O), 61 double-mode (F/FO).
In the second part of the OGLE-III Catalog of Variable Stars \citet{sos08b} have presented 197 Type\,II Cepheids and 83 ACs in the LMC. The ${V}$-band apparent visual magnitudes were transformed to the absolute visual magnitudes M$_{V}$ using the distance modulus of 18.53$\pm$0.02 given by \citet{mor12}. The Period-Luminosity relations have been determined from a linear regression to the the absolute visual magnitudes and log (P) values for the types of Cepheids in the form 

\begin{equation}
M_{V} = a log P + b 
\end{equation}
where M$_{V}$ is the visual absolute magnitude and P is the pulsation period in days. The coefficients are presented in Table 9 with their uncertainties. In the last two columns the correlation coefficients $R^2 $ and number of the stars used $N$ are also given. The Period-Luminosity relations for Cepheid variables in the LMC are shown in Fig.\,9. The individual FO and F pulsators of ACs are also plotted in the figure. The FO pulsators are brihter about 0.7 mag than the F pulsators at the fixed pulsation period. They lie between the FO and F pulsators of the Classical Cepheids (Type\,I Cepheids).  However, F pulsators of ACs are located between the F pulsators of the Type\,I Cepheids and Type\,II Cepheids in the P-L diagram in contrary to the suggestion by \citet{mar04} and \citet{cap04}. They propose that ACs and Type\,I Cepheids define a common region in the P-L plane, confirming the suggestion of \citet{dol03} and \citet{cor03} that the ACs are natural extension of the Type\,I Cepheids to lower metal contents and smaller mass. The pulsating stars in the binary systems are also plotted in the same diagram. All of them seem to be located at the same region covered by the F pulsators of the ACs. Absolute visual magnitudes, colors and pulsation periods are known for only six Type\,II binary Cepheids. The binarity of the distant cepheid QQ \,Per is doubtful \citep{wal08}.

\begin{table}
\scriptsize
\caption{The Period-Luminosity relations for the Type\,I Cepheids, Type\,II Cepheids and Anomalous Cepheids in the LMC. The formal uncertainties on the coefficients as well as the number of stars used and correlation coefficients of the linear fits are also given.}
\setlength{\tabcolsep}{1.8pt} 
\begin{tabular}{lcccc}
\hline
Absolute   & { $a $} & $b$& $R^2 $ & $N$                    \\
magnitude& (mag dex$^{-1}$) &(mag) &  &                    \\
\hline	
M$_{V}$(Cep\,I, FO)							&-3.148$\pm$0.036 					&-1.462$\pm$0.013	&0.86	  &1228		    \\
M$_{V}$(Cep\,I, F)	 						&-2.775$\pm$0.032					&0.953$\pm$0.021			&0.80        &1831			 \\
M$_{V}$(Cep\,II)							&-2.064$\pm$0.086					&0.409$\pm$0.086			&0.74       &202		    \\
M$_{V}$(AC\,, FO)	 						&-2.912$\pm$0.384					&-1.066$\pm$0.092			&0.76   	&20		 \\
M$_{V}$(AC\,, F)  	 						&-2.944$\pm$0.236					&-0.381$\pm$0.035			&0.73        &61		     \\
\hline
\end{tabular}
\end{table}

\begin{figure*}
\center
\includegraphics[width=16cm,angle=0]{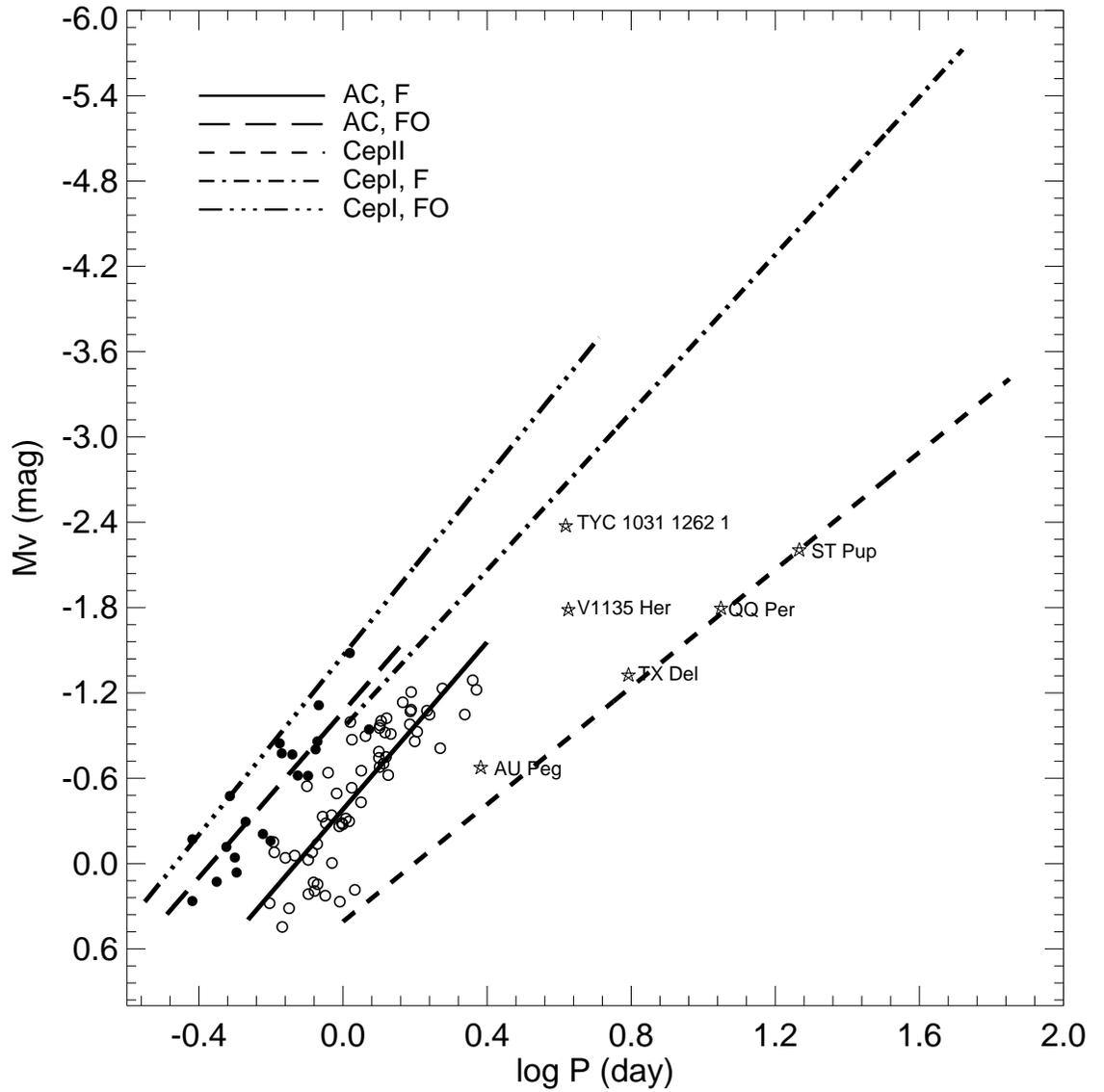}
\caption{The Period-Luminosity relations for the Cepheids in the LMC. Circles and dots represent fundamental and first-overtone Anomalous Cepheids. The star symbols represent binary Type II Cepheids, including 
V1135 \,Her. }
\end{figure*}

Individual masses of ACs could be estimated from the period-mass-amplitude (PMA) and period-luminosity-color relations (PLC). Using the equations given by \citet{mar04} for the F pulsators of ACs  we estimate a mass of 1.3 M$_{\odot}$ from the PMA relation and 1.8 M$_{\odot}$ from PLC relation. However, the relation for the FO pulsators gives a mass of 1.4 M$_{\odot}$. They seem to be in agreement within 3$\sigma$ with that found from our analysis . With a mass of about 1.461 M$_{\odot}$, luminosity of 517 L$_{\odot}$ and radius of 27 R$_{\odot}$ the pulsating primary component of V1135 \,Her cannot be a Type\,II Cepheid. If it were a Type\,I Cepheid its mass would be greater than 3.5 M$_{\odot}$ according to the models given by \citet{bon00} for Z=0.02 and even for Z=0.004. Recently \citet{fio12} suggested that ACs are metal-poor stars with the mean mass of 1.2$\pm$0.2  M$_{\odot}$. The mass, luminosity, kinematic properties and location of V1135 \,Her  in the C-M and P-L diagrams fulfil most of the properties of the Anomalous Cepheids. However, its pulsation period of 4.22 days is longer than that of the single ACs. The ACs in the LMC have pulsation periods generally shorter than 2.5 days.

Comparison of its location in the Hertzsprung-Russell diagram (HR), constructed for the metal-poor cepheids by \citet{gin85}, their Fig.\,1, indicates that V1135\,Her is in the instability strip of the Type\,II Cepheids, about 12 times more luminous than the RR \,Lyrae stars. It seems to have the same luminosity and effective temperature with the Type\,II Cepheids having pulsation periods of about 10 days, closer to 
the middle of the IS. When we compare with the solar composition models, for example \citet{bar98}, the primary star appears to be an evolved 5 M$_{\odot}$ star and the secondary is consistent with an evolved 3 M$_{\odot}$ star.

ACs are found in many nearby Local Group dwarf galaxies but rare in globular clusters. There is a general consensus that ACs are central helium burning stars with a mass about 1.2 M$_{\odot}$ which is too large with respect to Type\,II Cepheids, too small to Type\,I Cepheids. The origin of their progenitors is still unclear. There are two suggestions for their origin: a) the evolution of single, metal-poor star with mass 
$<$ 2 M$_{\odot}$ (\citet{bon97}, \citet{fio06}), and b) the  evolution of coalescent binary systems of metal poor stars (\citet{fio12}, \citet{sil09}). If the former is true they should be younger than the majority of stars in the globular clusters or dwarf spheroidals in which they are located (\citet{mat95}, \citet{cap99}). The evolutionary histories of V1135\,Her and other close binary systems having ACs are very uncertain.

\section{Conclusions }
We have obtained BVR light curves and radial velocities of the eclipsing binary system V1135\,Her of which the more massive star is a Cepheid. The astrophysical parameters, mass, radius, effective temperature and luminosity  of the component stars were obtained directly by analysing the light curves and radial velocities. The primary star was classified as a metal-poor Type\,II Cepheids, an analogue of TYC\,1031\,1262\,1. 
With a mass of 1.46 M$_{\odot}$ and radius of 27 R$_{\odot}$ its structure and evolution are different from those of the RR\,Lyrae and Type\,II Cepheids. Location of the more massive star on the C-M and P-L diagrams indicates that it may belong to metal-poor ACs. We could not determine the galactic velocity components due to very large uncertainties on the proper motions given in the literature. The distance from the Sun and galactic plane are found to be about 7500 and 1300 pc, respectively. All these properties and the asymmetric pulsating light curve lead us to classify the Cepheid as an Anomalous Cepheid. However, the pulsation period is longer, as in the other Type\,II Cepheid binaries, than those well-known ACs in the LMC. 

The pulsating primary star almost fills its corresponding Roche lobe while the less massive star fills only 30 per cent. Therefore, mass loss or transfer from the pulsating component to its companion  has taken place. Both components are located between giant and supergiant luminosity classes and the separation is only 1.6 times the sum of their radii. The companion should unexpectedly affect both pulsation and evolution of the Cepheid.  Most of the ACs are faint stars and have  orbital periods between 40 and 410 days, therefore, measurements of the radial velocities ACs in binary systems are very difficult. In addition the velocity variations due to the orbital motion may be affected by the pulsation depending on the inclination of the orbits. In the case of V1135\,Her the orbital inclination is sufficiently large to separate the orbital velocities of both components. If mass loss or transfer from the pulsating star has taken place, the orbital period of the system would be changed, which can be revealed by the observations to be made in coming years. This will yield us some clues about the further evolution of the ACs in binary systems.

\section*{Acknowledgment}
We thank to T\"{U}B{\.I}TAK National Observatory (TUG) for a partial support in using RTT150, T100 and T60 telescopes with project numbers 10ARTT150-483-0, 11ARTT150-123-0, 10CT100-101, 12BRTT150-338-1 and 10CT60-72. 
This study supported by the Turkish Scientific and Technology Council under project number 112T016 and 112T263.
We also thank to the staff of the Bak{\i}rl{\i}tepe observing station for their warm hospitality. 
The following internet-based resources were used in research for this paper: the NASA Astrophysics Data System; the SIMBAD database operated at CDS, Strasbourg, France; T\"{U}B\.{I}TAK 
ULAKB{\.I}M S\"{u}reli Yay{\i}nlar Katalo\v{g}u-TURKEY; and the ar$\chi$iv scientific paper preprint service operated by Cornell University. We are grateful to the anonymous referee, whose comments and suggestions helped to improve this paper.

\end{document}